\documentclass[aps,prb,showpacs]{revtex4}
\usepackage{graphicx}

\newcommand{\be}{\begin{equation}}
\newcommand{\ee}{\end{equation}}
\newcommand{\bea}{\begin{eqnarray}}
\newcommand{\eea}{\end{eqnarray}}
\newcommand{\bean}{\begin{eqnarray*}}
\newcommand{\eean}{\end{eqnarray*}}

\begin{document}




\title{Magnetic field effects and renormalization of the long-range Coulomb interaction in
Carbon Nanotubes}
\author{S. Bellucci $^a$ and P. Onorato $^a$ $^b$ \\}
\address{
        $^a$INFN, Laboratori Nazionali di Frascati,
        P.O. Box 13, 00044 Frascati, Italy. \\
        $^b$Dipartimento di Scienze Fisiche,
        Universit\`{a} di Roma Tre, Via della Vasca Navale 84,
00146 Roma, Italy} 
\date{\today}
\begin{abstract}
We develop two theoretical approaches for dealing with the low-energy
effects of the repulsive interaction in one-dimensional electron systems.
Renormalization Group methods allow us to study the low-energy
behavior of the unscreened  interaction between currents of
well-defined chirality in a strictly one-dimensional electron system. A
dimensional regularization approach is useful, when dealing with the
low-energy effects of the long-range Coulomb interaction. This
method allows us to avoid the infrared singularities arising from
the long-range Coulomb interaction at $D = 1$. We can also compare
these approaches with  the Luttinger model, in order to analyze the
effects of the short range term in the interaction.

Thanks to these  methods, we are able  to  discuss the effects of a
strong magnetic field $B$ in quasi one-dimensional electron systems, by focusing
our attention on Carbon Nanotubes. Our results imply a variation
with $B$ in the value of the critical exponent $\alpha$  for the
tunneling density of states, which is in fair agreement with that
observed in a recent transport experiment involving carbon
nanotubes. The dimensional regularization allows us to
predict the disappearance of the Luttinger liquid, when the
magnetic field increases, with the formation of a chiral liquid
with $\alpha=0$.

\end{abstract}


\maketitle
\section{Introduction}
In the last 20 years progresses in semiconductor device
fabrication and carbon technology allowed for the construction of
several new devices at the nanometric scale and many novel
transport phenomena have been revealed in mesoscopic
low-dimensional structures. The limits of further miniaturization
(predicted by Moore's law) have increased the research toward the
development of  electronics at the nanoscale and new efforts of
scientists have been stimulated by the progress in carbon and
semiconductor technology, aimed at building a nanoelectronics
\cite{1n,2n}. One-dimensional (1D) nanodevices, such as carbon
nanotubes (CNs) and semiconductor Quantum Wires (QWs), are the
building blocks of this new kind of electronics\cite{noiras}, and
recent experiments have revealed that they are also excellent
systems for the investigation of electronic transport in one
dimension (for other low
dimensional semiconductor devices, such as Quantum Dots, see e.g.\cite{noidot}).

Semiconductor QWs are quasi 1D devices  (having a width smaller than $1000
\AA$\cite{thor} and a length of some microns) made from a two-dimensional
electron gas (2DEG) created in heterostructures
between different thin semiconducting layers (typically
$GaAs:AlGaAs$) \cite{thor}. An
ideal Single Wall CN (SWCN)  is a hexagonal network of carbon
atoms (graphene sheet) that has been rolled up, in order to make a cylinder
with a radius about $1 nm$ and a length about $1 \mu m$. The unique
electronic properties of CNs are due to their diameter and chiral
angle (helicity)\cite{3n}. Multi Wall CNs (MWCNs), instead, are
made by several (typically 10) concentrically arranged graphene
sheets with a radius about $5 nm$ and a length about $1/100 \mu m$.

\
Electron transport in 1D devices  attracts considerable interest
 because of the fundamental importance of the
electron-electron (e-e) interaction in 1D systems: the e-e
interaction in a 1D system is expected to lead to the formation of
a so-called Tomonaga-Luttinger (TL) liquid with properties very
different from those of the non-interacting Fermi
gas\cite{egg,TL,TLreviwew}. It is well known that the TL liquid
behavior describes the regime with absence of electron
quasiparticles that is characteristic of 1D electron systems with
dominant repulsive interactions \cite{4n,5n,6n}.
The interest in TL liquids increased in recent years because of
the progresses in the experimental research about
CNs\cite{cnts,tubes} and semiconductor QWs \cite{tarucha,yacoby}.
 \ In these systems, the fact of having
 low-energy linear branches at the Fermi level introduces a
number of different scattering channels, depending on the location
of the electron modes near the Fermi points. It has been shown,
however, that processes which change the chirality of the modes,
as well as processes with large momentum-transfer (known as
backscattering and Umklapp processes), are largely subdominant,
with respect to those between currents of like chirality (known as
forward scattering processes)\cite{8n,9n,egepj}. Therefore CNs
and QWs should fall into the Luttinger liquid universality
class \cite{9n,egepj}. So, most experiments concentrated on
the power-law behavior of the tunneling density of states (DOS),
supporting that expectation.

\

From the theoretical point of view, a relevant question is the
determination of the effects of the long-range Coulomb interaction
in CNs. It is known  that the Coulomb interaction is not screened
in one spatial dimension \cite{13n,14n,noiprb}, although in
several analyses carried out for such systems, the e-e interaction
is taken actually as short-range (TL model). As we discussed in
previous papers\cite{npb}, the effects of the long-range Coulomb
interaction have been shown to lead in general to unconventional
electronic properties\cite{18n} and also to be responsible for a
strong attenuation of the quasiparticle weight in
graphite\cite{19n}.

\


In this paper we show how the
presence of a transverse  magnetic field modifies the role played by
the e-e interaction in a 1D electron system, also taking
into account its long range effects. The focal point is the
rescaling of all repulsive terms in the interaction between
electrons, due to the competition between the edge localization of
the electrons and the reduction of the magnetic length.
 Theoretically, it is predicted that a perpendicular magnetic
field modifies the DOS of a nanotube \cite{[33]}, leading to the
Landau level formation that  was observed in a MWNT
single-electron transistor\cite{[34]}.


The effects of a transverse magnetic field acting on MWNTs were
also investigated in the last few years: Kanda et al.\cite{kprl}
examined the dependence of the conductance $G$ on perpendicular magnetic fields.
They found that the exponent $\alpha$  depends significantly on the
 magnetic field and, in
most cases, $G$ is smaller for higher magnetic fields.
In particular, they showed that $\alpha$ is reduced from a value of
$0.34$ to a value $0.11$ for a magnetic field ranging from $0$ to
$4$ T (and from a value of $0.06$ to a value $0.005$ for a
different value of the gate voltage $V_g$).


Recently we discussed
the effects of a transverse magnetic field in QWs\cite{noimf} and
large radius CNs\cite{noimf1}.
In ref.\cite{noimf} we discussed the effects of a strong magnetic
field in QWs by focusing on the case of a very short range e-e
interaction. The presence of a magnetic field
produces  a strong reduction of the backward scattering due to the
edge localization of the electrons. This phenomenon can be easily
explained in terms of the Lorentz force which localizes the
opposite current at opposite edges of the device. The low-energy
behavior of Luttinger liquids is dramatically affected by
impurities which  can modify the conductance in the wire. In
ref.\cite{noimf} we also showed that the backward scattering
reduction and the rescaling of the e-e interaction
could favor the weak potential limit (strong tunneling), by raising
the temperature at which the wire becomes a perfect insulator
($G=0$).

In a previous paper\cite{24n}, we developed an analytic
continuation in the number D of dimensions, in order to accomplish the
renormalization of the long-range Coulomb interaction at
$D\rightarrow1$. The attenuation of the electron quasiparticles becomes
increasingly strong as $D\rightarrow1$, leading to an effective
power-law behavior of
 the tunneling DOS. In this way, we were able
to predict a lower bound of the corresponding exponent, which turned
out to be very close to the value measured in experimental
observations of the tunneling conductance for MWNTs\cite{22n}.
More recently\cite{npb} we introduced the effect of the number of
subbands that contribute to the low-energy properties of CNs.
This issue was relevant for the investigation of the
nanotubes of large radius that are present in MWNTs, which are
usually doped and may have a large number of subbands crossing the
Fermi level \cite{25an}.


In this paper, we focus on the presence of the magnetic field and the
long range electron repulsion in CNs, by using two different
approaches which allow us to calculate the different values of the
critical exponent $\alpha $ measured for the tunneling DOS.

We also propose  different models for the e-e
interaction, i.e. an unscreened Coulomb interaction in two dimensions and
a generalized Coulomb interaction in arbitrary dimensions, which
allows us to implement the dimensional crossover approach.

With our calculations we explain the observed reduction of the
critical exponent $\alpha$ corresponding to the tunneling DOS for
a quasi 1D electron systems. In particular, this approach allows
us to fit the recently measured behavior of MWNTs under the effect
of a strong magnetic field.

\
 \

\section{Band structure of graphene and single particle hamiltonian for CNs}

The band structure of CNs can be obtained by the
technique of projecting the band dispersion of a two-dimensional (2D) graphite layer
into the 1D longitudinal dimension of the nanotube. The 2D band
dispersion of graphene\cite{25n} consists of an upper and a lower
branch that only touch each other at the corners of the hexagonal Brillouin
zone.

The 2D layers in graphite have a honeycomb structure with a simple
hexagonal Bravais lattice and two carbon atoms in each primitive
cell, so the tight-binding calculation for the honeycomb lattice
gives the known bandstructure of graphite
\begin{equation}
E ({ \bf k})= \pm \gamma \sqrt{1+4\cos^2(\frac{\sqrt{3} }{2}k_x)
+4{\cos(\frac{\sqrt{3}k_x}{2})}\cos(\frac{{3}k_y}{2})}.
\end{equation}
After the definition of the boundary condition (i.e. the wrapping
vector $\overrightarrow{w}=(m_w,n_w)$), it is easy to obtain the
bandstructure of the CN that can be approximated by the formula
$$
\varepsilon_0({m},\overrightarrow{w}, k)\approx \pm  \frac{v_F
\hbar}{R} \sqrt{\left(\frac{m_w-n_w+3m}{N_b}\right)^2 +R^2 \left(k
\pm K_s\right)^2},
$$
where $R$ is the radius of the tube, connected to the value of
$N_b$ in a simple way  $R\approx N_b \sqrt{3} a/(2\pi)$, $a$
denotes the honeycomb lattice constant ($a/\sqrt{3}=a_0 = 1.42
\AA$),  $K_s=\frac{2\,\pi }{3\,a{\sqrt{3}}}$ and $v_F$ is the
Fermi velocity ($v_F\approx 10^6 m/s$).

For a metallic CN (the armchair one with $m_w=n_w$) we obtain that
the energy vanishes for two different values of the longitudinal
momentum $ \varepsilon_0(\pm K_s)=0$. The dispersion law
$\varepsilon_0({m}, k)$ in the case of undoped metallic
nanotubes is quite linear near the crossing values $\pm K_s$. The
fact of having four low-energy linear branches at the Fermi level
introduces a number of different scattering channels, depending on
the location of the electron modes near the Fermi points\cite{7n}.

Now we assume that the eigenfunctions corresponding to the
dispersion law written above are ${\Phi}_{m,k}(\varphi,y)$, so that
$$ H_0
{\Phi}_{m,k}(\varphi,y)=\varepsilon_0(m,\overrightarrow{w},k){\Phi}_{m,k}(\varphi,y).$$
Next we approximate $ \Phi_{m,k}(\varphi,y)\approx
\frac{e^{\frac{i m \varphi}{2 \pi R}}e^{i k y}}{2 \pi L^{1/2} }$.

Now we consider a tube with the axis along the $y$ direction,
under the action of a magnetic field along the $z$ direction, and
we choose the gauge, so that the system has a symmetry along the
${y}$ direction, ${\bf A}=(0,Bx,0)$. So we can write the magnetic
term of the Hamiltonian as
\begin{equation}\label{h1}
H_1=H-H_0 = - \omega_c R \cos(\varphi) p_y + \frac{m_e\omega_c^2
R^2}{2} \cos^2(\varphi) ,
\end{equation}
where  $\omega_c=\frac{eB}{mc}$ is the cyclotron frequency. The
term  $H_1$ can be taken as a perturbation, and it gives
corrections to the energy to the second order in $\omega_c$. If we
introduce the magnetic length $\ell_\omega=\sqrt{\hbar/(m
\omega_c)}$ and a constant
$\gamma=\frac{R^2}{\ell_\omega^2}\frac{\hbar k}{m v_F}$, we can
write
$$
\delta \varepsilon_{0,k}=\pm\left(-\frac{ \hbar v_F k^2
\gamma^2}{R}+ \frac{m_e\omega_c^2 R^2}{4 }\right) \; \; \; \; \;
\; \; \;
 \widetilde{v}_F \approx  v_F \left(1 -\frac{2\gamma^2}{3 N_b} \right),
$$
for the correction to the energy and to the Fermi velocity  (here
shown for the lowest subband $m=0$). For the lowest subband
($m=0$) the perturbed eigenfunctions are given by \bea\label{wft}
 \widetilde{\Phi}_{0,\pm k}(\varphi,y)=N_0
\left(1\pm 2{\gamma}  \cos(\varphi)\right)e^{i k y}\equiv
u_0(\varphi,k) \frac{e^{i k y}}{\sqrt{2\pi L}}, \eea
where  $N_0$ is  the normalization constant $
N_0=\sqrt{\frac{1}{1+ {\gamma}^2 }}$.

\section{Interaction Models}
As we discussed previously, in this paper we limit ourselves to
the one channel model ($n=n'=0$), i.e. to the magnetic field
dependent effective potential
\begin{eqnarray}\label{intpot}
\nonumber U_{{\bf k,p,q}}({\bf r-r'},\omega_c) =
\int_{0}^{2\pi}\int_{0}^{2\pi} d\varphi d\varphi' U(|{\bf
\widetilde{r}}-{\bf \widetilde{r}}'| ) u_{0}\left(\varphi,
k\right)u_{0}\left(\varphi,p\right) u_{0}\left(\varphi',
(k+q)\right) u_{0}\left(\varphi',(p-q)\right) \;.
\end{eqnarray}
Here ${\bf r}$ is a vector in the $D$ dimensional space and  ${\bf
\widetilde{r}}$ is a vector in  $D+1$ dimension. Next we need
$U_0({\bf q})$ which  corresponds to the Fourier transform of the
Coulomb potential in dimension $D$.

\

Following the procedure shown above, we can start from the
unscreened Coulomb interaction in two dimensions, in agreement
with Egger and Gogolin\cite{egepj}  \bea\label{int2} U({\bf
r}-{\bf r'})=\frac{c_0}{\sqrt{(y-y')^2+4 R^2
\sin^2(\frac{\varphi-\varphi'}{2})}},  \eea and, after calculating
the effective $1D$ potential $U_{{\bf k,p,q}}({ y-y'},\omega_c)$,
we obtain a formula for the forward scattering term as
\bea\label{vq} U_0(q,\omega_c)=\frac{c_0}{\sqrt{2}\left( 1 + 2
\,{\gamma }^2
        \right)^2}\left[K_0(\frac{qR}{2})I_0(\frac{qR}{2}) +{\gamma}^2
        \left(2G_1(\frac{q^2R^2}{4})+G_2(\frac{q^2R^2}{4})\right)\right],
\eea where  $K_n(q)$  gives the modified Bessel function of the
second kind, $I_n(q)$ gives gives the modified Bessel function of
the first kind
 and $G_i$ are expressed in terms of the $MeijerG$ functions.

Some details about this calculation are shown in Appendix A, where
we also calculate the backward scattering term. As we show in
Fig.(1,left), this term of the coupling is strongly suppressed by
the presence of the transverse  magnetic field.
\begin{figure}
\includegraphics*[width=1.00\linewidth]{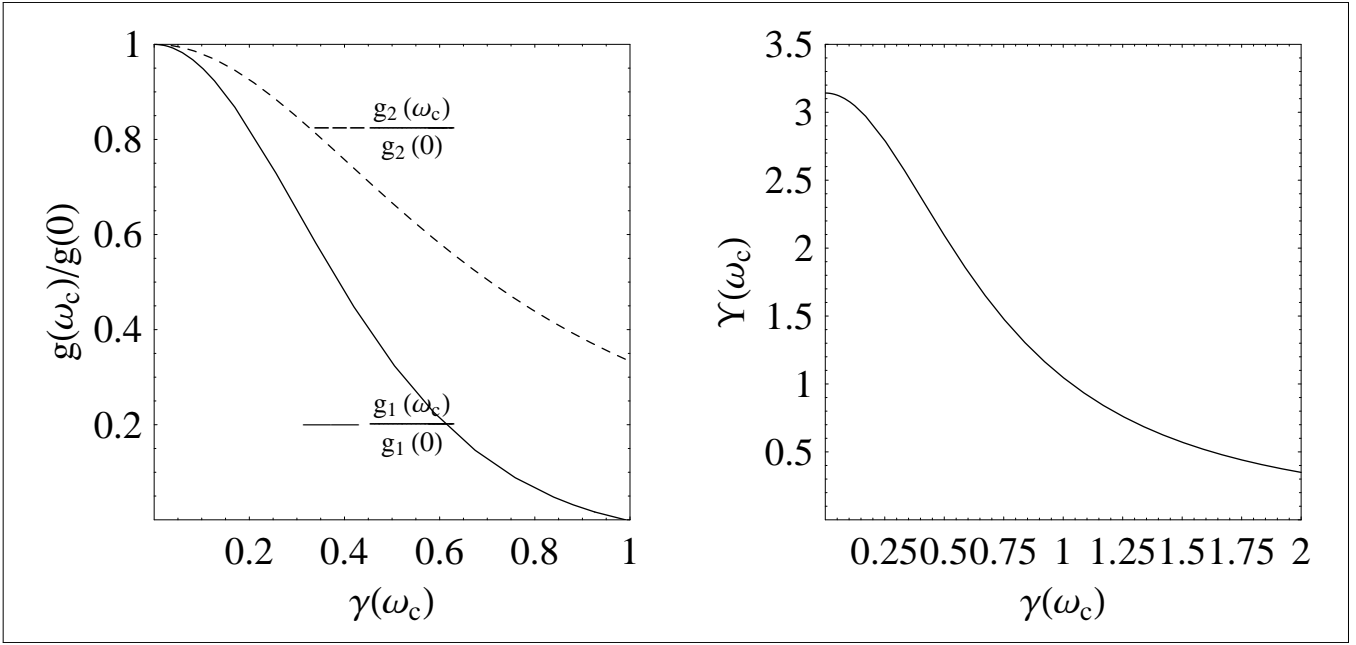}
 \caption{(Left) Backscattering suppression for the unscreened 1D Coulomb
 interaction in a CN.
Here we introduce the  parameter as $\gamma(\omega_c)\propto
\omega_c$.
We can observe how the interaction coupling reduction
becomes less strong, as the range of the interaction increases. (Right)
Correcting Factor $\Upsilon$ for the Coulomb interaction in
arbitrary dimensions. }
\end{figure}

\

Because we want  develop a dimensional regularization approach,
following the calculations in ref.\cite{19n}, in order to analyze
the low energy effects of the divergent long-range Coulomb
interaction in one dimension, we have to introduce the interaction potential
in arbitrary dimensions
 \bea \label{ud}
 U_D({\bf r}-{\bf r'})=
 \int_{0}^{2\pi}\int_{0}^{2\pi} d\varphi d\varphi' \frac{c_D}{4\pi^2|{\bf \widetilde{r}}-{\bf \widetilde{r}'}|}
 u_{0}\left(\varphi,
k\right)u_{0}\left(\varphi,p\right) u_{0}\left(\varphi',
(k+q)\right) u_{0}\left(\varphi',(p-q)\right) =
        \frac{c_D}{|{\bf r}-{\bf r'}|}\Upsilon_2(\omega_c).
 \eea

Here ${\bf r}$ is a vector in the $D$ dimensional space and  ${\bf
\widetilde{r}}$ is a vector in  $D+1$ dimensions.
\

As it is known, the Coulomb potential $1/|{\bf r}|$ can be
represented in three spatial dimensions as the Fourier transform
of the propagator $1/{\bf k}^2$
\begin{equation}\label{ur}
\frac{1}{|{\bf r}|} =  \int \frac{d^3 k}{(2\pi )^3}
      e^{i {\bf k \cdot r}   }
       \;   \frac{1}{ {\bf k}^2 }.
\end{equation}

If the interaction is projected onto one spatial dimension, by
integrating for instance the modes in the transverse
dimensions, then the Fourier transform has the usual logarithmic
dependence on the momentum\cite{14n}. We choose instead to
integrate formally a number $3-D$ of dimensions, so that the
long-range potential gets the representation
\begin{equation}\label{ux}
\frac{1}{|x|} =  \int \frac{d^D k}{(2\pi )^D} e^{ikx}
       \;   \frac{c(D)}{|{\bf k}|^{D-1}},
\end{equation}
where $c(D) = \Gamma ((D-1)/2)/(2\sqrt{\pi})^{3-D}$.

In order to  introduce the effects of the magnetic field in a
model of interaction at arbitrary dimensions, we start from the
three equations above. We can rescale the constant  $c(D)$ by
considering the factor $\Upsilon_2(\omega_c)$, which contains the
effects of the edge
states localization, so that
\bea \label{intd}U(p)=
\Upsilon_2(\omega_c)\frac{c_0(D)}{|{\bf p}|^{D-1}}\equiv
\frac{c(D,\omega_c)}{|{\bf p}|^{D-1}}. \eea The factor
$\Upsilon_2(\omega_c)$ is calculated in appendix B and is plotted
in Fig.(1,right), where we show the dependence of this factor on the
magnetic field.

\section{RG solution for D=1}

In a previous paper  \cite{npb}, we have developed a RG approach,
in order to regularize the infrared
singularity of the long-range Coulomb interaction.
 Our aim was to
find the effective interaction between the low-energy modes of
CNs, which have quite linear branches near the top of the subbands
($K_S$).
 We start
then with the Hamiltonian
\begin{eqnarray}
H  =
    \int_0^{\Lambda }   \frac{d p }{(2\pi )}
   \psi^{+} ({p})
   \varepsilon(p)
            \psi ({p})
   {  +    \int_0^{\Lambda }  \frac{d p }{(2\pi )}\;
   \rho ({p})   \;   U_0(p) \;
          \rho (-{ p})  \;\;\;\;\;\; }
\label{ham0}
\end{eqnarray}
where  $\rho ({p})$ are density operators made of the electron
modes $\psi ({p})$, and $ U_0(p) $ corresponds to the Fourier
transform of the 1D interaction potential.

\ Here we follow the calculations published in ref.\cite{14n}. The
main difference is the introduction of the non singular model of
the magnetic field dependent interaction shown above, instead of
the usual Coulomb long range interaction.

In writing eq.(\ref{ham0}), we have neglected backscattering
processes that connect the two branches of the dispersion
relation. This is justified, in a first approximation, as for the
Coulomb interaction the processes with small momentum transfer
have a much larger strength than those with momentum transfer $\sim 2
k_F$. The backscattering processes give rise, however, to a
marginal interaction and, as we have shown in the previous section,
they are strongly reduced by the magnetic field.

 \
 {The one-loop polarizability $\Pi_0(k,\omega_k)$ is given by the
sum of particle-hole contributions within each branch {
\begin{equation}
\Pi_0({\bf k}, \omega_k) =  \frac{v_F {\bf k}^2}
 { | v_F^2 {\bf k}^2 - \omega_k^2 | }\; \;.
\label{pol0}
\end{equation}

The effective interaction is found by the Dyson equation
\begin{eqnarray}
U_{eff} ({ k}, \omega_k)  & = & \frac{U_0({ k})}{1-U_0({
k})\Pi_0({ k}, \omega_k)}, \label{ueff}
\end{eqnarray}
so that the self-energy follows: $\Sigma_{eff}=G_0 U_{eff}=G_0
U_{eff}=\frac{G_0 U_0}{1-U_0\Pi_0} $.

 In the spirit of the GW approximation, we consider
$ {v}_F$ as a free parameter that has to match the Fermi velocity
in the fermion propagator after self-energy corrections.

The polarization gives the effective interaction $U_{eff}$ as in
eq.(\ref{ueff}) which incorporates the effect of plasmons in the
model. We compute the electron self-energy by replacing the
Coulomb potential by the effective  interaction calculated
starting with our model of interaction in eq.(\ref{vq})
\begin{equation}
i \Sigma (k, i \omega_k) = i \frac{e^2}{2\pi } \int^{E_c}_{-E_c}
\frac{dp}{2\pi }
 \int^{+\infty}_{-\infty}
\frac{d\omega_p}{2\pi } \frac{1}{i (\omega_p + \omega_k) - v_F (p
+ k) } \frac{U_0(p) }{1 -  \frac{e^2}{\pi }
 \frac{ {v}_F p^2}{ {v}_F^2
p^2 + \omega_p^2 } U_0(p) }. \label{selfe0}
\end{equation}

{ Below we  show that this approximation reproduces the exact
anomalous dimension of the electron field in the Luttinger model
with a conventional short-range interaction}.

The only contributions in (\ref{selfe0}) depending on the
bandwidth cutoff are terms linear in $\omega_k$ and $k$.
 There is
no infrared catastrophe at $\omega_k \approx v_F k$, because of
the correction in the slope of the plasmon dispersion relation,
with respect to its bare value $v_F$. The result that we get for
the renormalized electron propagator is
\begin{eqnarray}
G^{-1}(k,\omega_k) & = &  Z^{-1}_{\Psi} \; (\omega_k - v_F
  k)  - \Sigma (k,\omega_k)  \nonumber   \\
& \approx & Z^{-1}_{\Psi} \; (\omega_k - v_F  k) + Z^{-1}_{\Psi}
\; (\omega_k - v_F k)  \int^{E_c}
 \frac{dp}{|p|}
\frac{(1 - f(p))^2 }{2 \sqrt{f(p)} \left( 1 + \sqrt{f(p)}\right)^2
} +\ldots , \label{gren}
\end{eqnarray}
where $f(p) \equiv 1 +U_0(p,\omega_c) /(2 \pi  {v}_F)  $ and
$Z^{1/2}_{\Psi}$ is the scale of the bare electron field compared
to that of the cutoff-independent electron field
\begin{equation}
\Psi_{bare}(E_c) = Z^{1/2}_{\Psi} \Psi \;.\;
\end{equation}

The first RG flow equations, obtained analogously to the more
general  eq. (\ref{zflow}) obtained below, becomes
\begin{eqnarray}
E_c \frac{d}{dE_c }\:  \log \: Z_{\Psi}(E_c)  & = & \frac{ \left(
1 - \sqrt{f(E_c)} \right)^2 }
    {8 \sqrt{f(E_c)}} .    \label{zflow2}
\end{eqnarray}
}

As it is known\cite{npb}, the critical exponent can be easily
obtained from the right side of  eq.(\ref{zflow2}) in the limit of
$\log(E_c)\rightarrow 0$.

 In the
case of a short range interaction, where $U(q)$ is a constant,
$g$, we can write
$$
\sqrt{f(q)}=\sqrt{1 + \frac{g}{ (2 \pi  {v}_F)}} =K.
$$
Hence, as it is clear from a comparison with refs.\cite{egepj} and \cite{sh}, we have
$$
 \alpha_Z=\frac{ \left(
1 - \sqrt{f(E_c)} \right)^2 }
    {8 \sqrt{f(E_c)}}   =\frac{1}{4} \left(K+\frac{1}{K}-2
    \right)\equiv (K^2-1) T_1(K).
$$
In the general case of a generic interaction we have to introduce
the infrared limit of the $\alpha$ function, as we will do in the next
section for the Coulomb repulsion. The critical exponent has the
form \bea\label{az0}
 \alpha_Z=\frac{ \left(
1 - \sqrt{f(q_c)} \right)^2 }
    {8 \sqrt{f(q_c)}},
\eea
 where $q_c$ has to be taken to be the natural infrared cutoff
$2 \pi /L$.

\section{Dimensional regularization near D=1}

In a previous paper  \cite{24n}, we have developed an analytic
continuation in the number of dimensions, in order to regularize
the infrared singularity of the long-range Coulomb interaction at
$D = 1$  \cite{ccm}. Our aim was to find the effective interaction
between the low-energy modes of CNs, which have quite linear
branches near the top of the subbands ($K_S$).
For this purpose,
we have dealt with the analytic continuation to a general dimension
$D$ of the linear dispersion around each Fermi point. We start
then with the Hamiltonian
\begin{eqnarray}
H & = &  v_F  \sum_{\alpha\sigma}
    \int_0^{\Lambda } d p |{\bf p}|^{D-1}
     \int  \frac{d\Omega }{(2\pi )^D} \;
   \psi_{\alpha\sigma}^{+} ({\bf p}) \;
    \mbox{\boldmath $\sigma
  \cdot $} {\bf p} \;
            \psi_{\alpha\sigma} ({\bf p})
   {  +   e^2 \int_0^{\Lambda } d p |{\bf p}|^{D-1}
     \int  \frac{d\Omega }{(2\pi )^D}  \;
   \rho ({\bf p})   \;   \frac{c(D)}{|{\bf p}|^{D-1}}  \;
          \rho (-{\bf p}),  \;\;\;\;\;\; }
\label{ham}
\end{eqnarray}
where the $\sigma_i $ matrices are defined formally by $ \{
\sigma_i , \sigma_j \} = 2\delta_{ij}$. Here $\rho ({\bf p})$ are
density operators made of the electron modes $\psi_{\alpha\sigma}
({\bf p})$, and $ c(D)/|{\bf p}|^{D-1} $ corresponds to the
Fourier transform of the Coulomb potential in dimension $D$. Its
usual logarithmic dependence on $|{\bf p}|$ at $D = 1$ is obtained
by taking the 1D limit with $c(D) = \Gamma
((D-1)/2)/(2\sqrt{\pi})^{3-D}$.

A self-consistent solution of the low-energy effective theory has
been found in\cite{14n} by determining the fixed-points of the RG
transformations implemented by the reduction of the cutoff
$\Lambda $.  In this section we adopt a Renormalization Group
theory with a dimensional crossover, starting from Anderson
suggestion\cite{and}  that the Luttinger model could be extended to
2D systems. The dimensional regularization approach
of ref.\cite{14n}, which we follow here, overcomes the problem of
introducing such an external parameter.

A phenomenological solution of the model was firstly obtained
\cite{14n}, carrying a dependence on the transverse scale needed
to define the 1D logarithmic potential, which led to
scale-dependent critical exponents and prevented a proper scaling
behavior of the model \cite{14n,wang}.


 Here
we assume that  the long-range Coulomb interaction  may lead
to the breakdown of the Fermi liquid behavior at any dimension
between $D=1$ and $D=2$, while the MWNT  description lies
between that of a pure 1D system and the 2D graphite layer. Then
we introduce  an analytic continuation in the number $D$ of
dimensions which allows us to carry out the calculations needed,
in order to accomplish the
renormalization of the long-range Coulomb interaction at
$D\rightarrow 1$.

In the vicinity of $D=1$, a crossover takes place to a behavior
with a sharp reduction of the electron quasiparticle weight and
the DOS displays an effective power-law behavior,
with an increasingly large exponent. For values of $D$ above the
crossover dimension, we have a clear signature of quasiparticles at
low energies and  the DOS approaches the well-known
behavior of the graphite layer.

As in the previous section, we start from the one-loop
polarizability $\Pi_0(k,\omega_k)$ given by the sum of
particle-hole contributions within each branch. Now it  is the
analytic continuation of the known result in eq.(\ref{pol0}),
which we take away from $D=1$, in order to carry out a consistent
regularization of the Coulomb interaction
 {
\begin{equation}
\Pi_0({\bf k}, \omega_k) = b(D) \frac{v_F^{2-D} {\bf k}^2}
 { | v_F^2 {\bf k}^2 - \omega_k^2 |^{(3-D)/2} }, \; \;
\label{pol}
\end{equation}
where $b(D) = \frac{2}{ \sqrt{\pi} } \frac{ \Gamma ( (D+1)/2 )^2
   \Gamma ( (3-D)/2 ) }{ (2\sqrt{\pi})^D \Gamma (D+1) }$.}
The effective interaction is found by the Dyson equation in
eq.(\ref{ueff}), so that the self-energy
$\Sigma_{eff}$ follows.
After dressing the interaction with the
polarization (\ref{pol}), the electron self-energy is given by the
expression
\begin{eqnarray}
\Sigma ({\bf k}, \omega_k)  & = &  - e^2 \int_0^{E_c /v_F }
     d p |{\bf p}|^{D-1}  \int \frac{d\Omega }{(2\pi )^D}
    \int \frac{d \omega_p}{2\pi }     \nonumber   \\
 \lefteqn{   G ({\bf k} - {\bf p}, \omega_k - \omega_p)
 \frac{-i}{ \frac{|{\bf p}|^{D-1}}{c(D)} + e^2  \Pi ({\bf p},
    \omega_p) }     }.
\label{selfe}
\end{eqnarray}
At general $D$, the self-energy (\ref{selfe}) shows a logarithmic
dependence on the cutoff at small frequency $\omega_k$ and small
momentum ${\bf k}$. This is the signature of the renormalization
of the electron field scale and the Fermi velocity. In the
low-energy theory with high-energy modes integrated out, the
electron propagator becomes
\begin{eqnarray}
\frac{1}{G}  & = &  \frac{1}{G_0} - \Sigma
     \approx  Z^{-1} ( \omega_k - v_F
  \mbox{\boldmath $\sigma \cdot$}{\bf k}) - Z^{-1} f(D)
 \sum_{n=0}^{\infty} (-1)^n g^{n+1}    \left(
   \frac{n(3-D)}{n(3-D)+2}  \omega_k    \right.     \nonumber   \\
 \lefteqn{ +  \left.  \left(1 - \frac{2}{D} \frac{n(3-D)+1}{n(3-D)+2}
   \right)   v_F \mbox{\boldmath $\sigma \cdot$} {\bf k}
      \right)  h_n (D)   \log (\Lambda ) ,  }
\label{prop}
\end{eqnarray}
where $g =(2b(D)c(D)e^2)/v_F$, $f(D) = \frac{1}{ 2^D \pi^{(D+1)/2}
\Gamma (D/2) b(D) }$ and $h_n (D) = \frac{ \Gamma (n(3-D)/2 + 1/2)
}
 { \Gamma (n(3-D)/2 + 1) }$.
The quantity $Z^{1/2}$ represents the scale of the bare electron
field compared to that of the renormalized electron field for
which $G$ is computed.

The effective coupling $g$ is a function of the cut off with an
initial value obtained carrying out an expansion near
$D=1$\cite{npb},
\bea \label{gd}
 g_0(D,\omega_c) =  c(D,\omega_c)  \frac{ e^2} {  \widetilde{v_F} }  \approx
  \Xi({\omega_c}) \frac{ e^2} {  \pi^{2}v_F} \frac{1}{D-1}=\Xi({\omega_c})  \frac{4 \widetilde{g}}{D-1}.
\eea
 The dimensionless factor
$\Xi({\omega_c})=v_F/\widetilde{v_F}\Upsilon_2({\omega_c})$
contains the scaling of the effective interaction with the
magnetic field, due to both the factor $\Upsilon_2({\omega_c})$  and
the scaling of the Fermi velocity $\widetilde{v_F}$.

 The renormalized
propagator $G$ must be cutoff-independent, as it leads to
observable quantities in the quantum theory. This condition is
enforced by fixing the dependence of the effective parameters $Z$
and $v_F$ on $\Lambda $, as more states are integrated out from
high-energy shells. We get the differential renormalization group
equations
\begin{eqnarray}
\Lambda \frac{d}{d \Lambda} \log Z (\Lambda )&=&-f(D)
\sum_{n=0}^{\infty}  {  \frac{n(3-D)(-g)^{n+1}}{n(3-D)+2} h_n
(D)}=-\gamma(g) \label{zflow}, \\\Lambda \frac{d}{d \Lambda}
g(\Lambda )&=&-f(D)\frac{2(D-1)}{D} g^2 \sum_{n=0}^{\infty} (-g)^n
{ \frac{(3-D)n+1}{(3-D)n+2} h_n (D)}=-\beta(g) \label{gflow} .
\label{vflow}
\end{eqnarray}
{For $D=1$ the function in the r.h.s. of eq.(\ref{gflow})
vanishes, so that the 1D model has formally a line of
fixed-points, as it happens in the case of short-range
interaction. In the crossover approach shown in this section, the
effective coupling $g$  is sent to strong coupling in the limit
$D\rightarrow 1$, and the behavior of the RG flow in this regime
remains to be checked.  We should also stress the dependence on
$D$ of the functions appearing in the RG equations, which shows
itself in the form of $D - 1$ and $D - 3$ factors, revealing that
these are the two critical dimensions, corresponding to a marginal
and a renormalizable theory, respectively.

 In the limit $D \rightarrow 1$  the
series in the r.h.s. can be also summed up , with the result that
the scaling equation in that limit corresponds to the one obtained in the
previous section by putting there $K=\sqrt{1+g_{eff}}$.

\subsection{RG scaling and low-energy density of states}
\begin{figure}
\includegraphics*[width=0.7\linewidth]{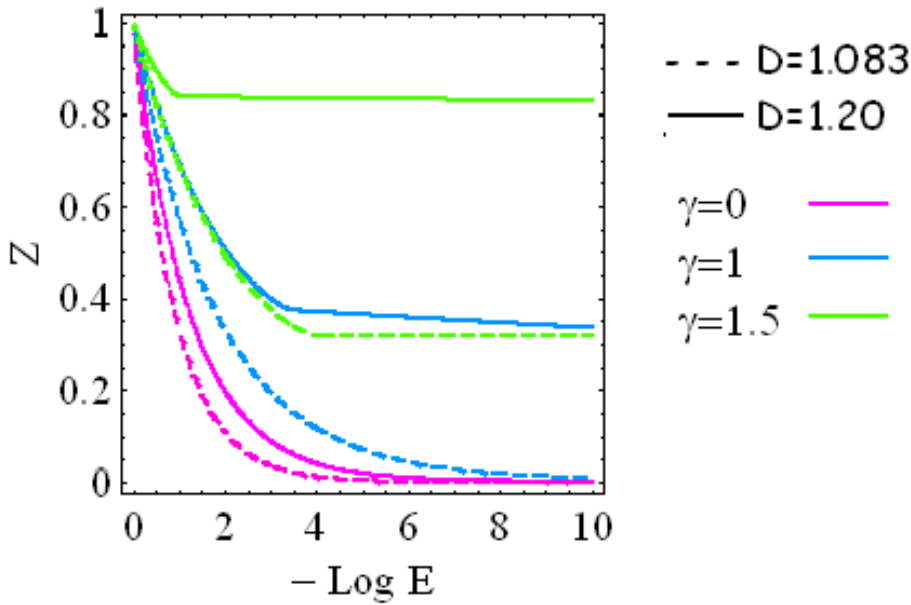}
 \caption{Energy
dependence of the quasiparticle weight Z at dimensions $D=1.2$ and
$D=1.083$ (dashed lines), for different values of $B$ parameterized
by $\gamma$ ($\gamma=1$ corresponds to a magnetic field of about $4$ T).
It is clear that, for a vanishing field, the quasiparticle weight is
renormalized to zero in both dimensions, while for a very strong
magnetic field, the Luttinger Liquid disappears at any dimensions
($\gamma=1.5$).
 }
\end{figure}
{As discussed in our previous papers, near $D=1$ we find a
crossover to a behavior with a sharp reduction of the
quasiparticle weight in the low-energy limit. This is displayed in
Fig.(2), where we have represented the electron field scale
Z. If the magnetic field increases (the dashed line corresponds to B
above 2 T), we have a clear signature of quasiparticles in the
nonzero value of Z at low energies, whereas for vanishing $B$ the
picture cannot be distinguished from that of a vanishing
quasiparticle weight.

We observe from the results in Fig.(2) that the quasi- particle
weight Z tends to have a flat behavior at high energies, for large
values of the magnetic field $B$, (this happens because of the
renormalization of the effective coupling to zero). This is in
contrast to the rapid decrease signaling the typical power-law
behavior, for small values of B.}

The dimensional crossover approach allows us  to
calculate the critical exponent also in this case of a
divergent interaction for $D \rightarrow 1$. In fact our target is
to compare theoretical results with measurements of the tunneling
DOS carried out in nanotubes when a strong magnetic
field acts on them. The DOS computed at dimensions
between $1$ and $2$ displays an effective power-law behavior
which is given by
$n( \varepsilon ) \sim Z( \varepsilon )
|\varepsilon |^{D-1}$,
for several dimensions approaching $D=1$. Then we introduce the
low-energy behavior of $Z( \varepsilon )$ in order to analyze the
linear dependence of $\log(n( \varepsilon ))$ on
$x=-\log(\Lambda)$
\begin{eqnarray} \label{ald} \log(n( \varepsilon )
)&\approx&  \log{Z( \varepsilon )}+ (D-1) \log(|\varepsilon
|)\approx  \left( \alpha_Z -(D-1) \right)x \equiv\alpha_D x. 
\end{eqnarray}

Here $\alpha_Z$ can be easily written starting from
eq.(\ref{zflow}), if we limit ourselves to a simple first order
expansion near $x = 0$ with ($ \log(Z) \sim \gamma (g_0 ) x$),
where $g_0$ is the initial value of the coupling (see
eq(\ref{gd})
\begin{eqnarray}\label{alphaz1}
  \alpha_Z \approx
  T_1(\sqrt{1+g_0})g_0 \frac{(3-D)f(D)(D+1)}{8}.
\end{eqnarray}

The analytic continuation in the number of dimensions  allows us to
avoid the infrared singularities that the long-range Coulomb
interaction produces at $D=1$, providing insight, at the same time,
about the fixed-points and universality classes of the theory in
the limit $D\rightarrow 1$. In order to compare our results with
experiments, as in ref.\cite{npb}, we can obtain a lower bound for
the exponent of the DOS by estimating the minimum of
the absolute value of $\alpha_D$, for dimensions ranging between
$D=1$ and $D=2$. The evaluation can be carried out starting from
eq.(\ref{ald}) and eq.(\ref{alphaz1}). We obtain a minimum value
for $|\alpha_D|$ as a function of $D$, when we introduce the
expression of $g_0(D,\omega_c)$ in eq.(\ref{gd}).
\begin{figure}
\includegraphics*[width=1\linewidth]{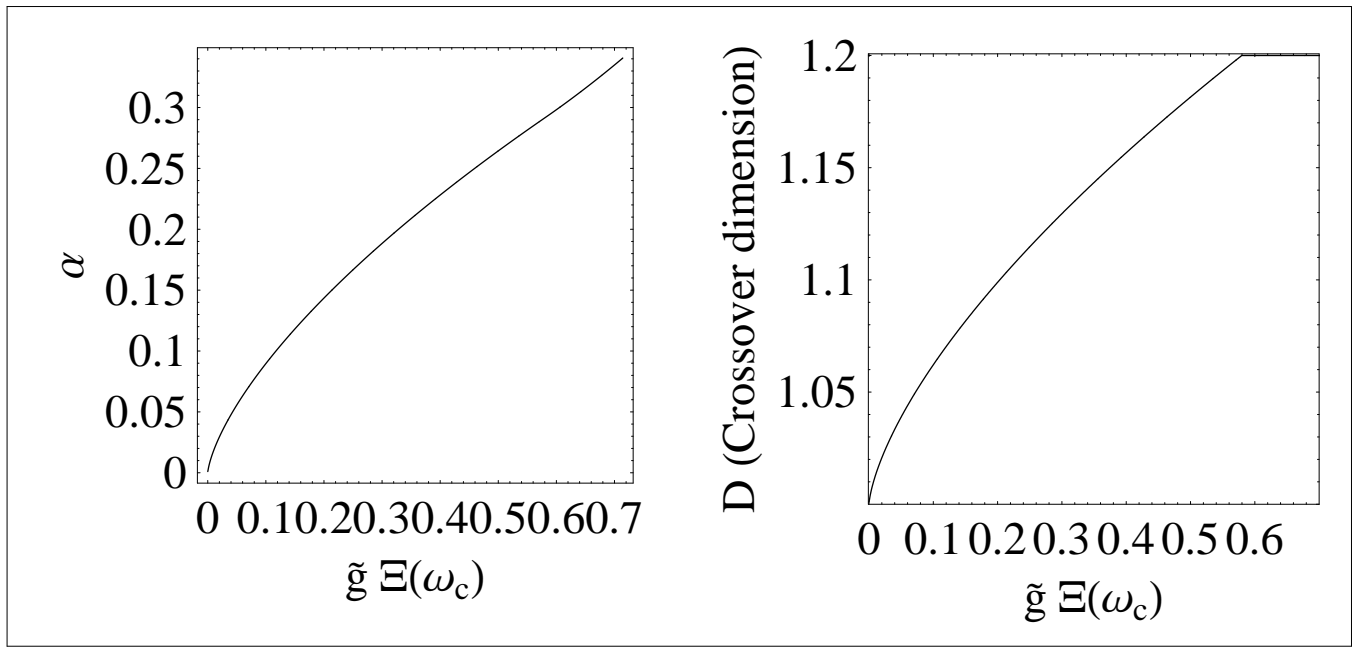}
 \caption{(Left) The critical exponent as a function of the effective coupling
  $g\Xi$ obtained with a numerical calculation. (Right) Crossover dimension as
   a function of the effective coupling. When $B$ increases, the effective coupling
   is strongly reduced and the crossover dimension approaches the value $1$ according
   to Fig.(2), where the stability of the quasiparticle weight is reported.   }
\end{figure}

 The resulting value of the critical exponent $\alpha$
is
$$
\alpha_0(\omega_c) \approx \left(
\Xi({\omega_c})\widetilde{g}\right)^{1/3}\approx
\Xi({\omega_c})^{1/3}\, \alpha(0)\longrightarrow\approx \left(
\Xi({\omega_c})\widetilde{g}\right)^{2/3}\approx
\Xi({\omega_c})^{2/3}\, \alpha(0)
$$
while in Fig.(3) we show, in the left panel,  the numerical values
of $\alpha$ and, in the right panel, the corresponding crossover
dimension. We can observe that the growth of the magnetic field,
corresponding to a reduction of the coupling $g \Xi$, reduces
strongly the crossover dimension, below which the quasiparticle
weight vanishes.

\section{Discussion}

 In this paper we have described two different
frameworks for dealing with the low-energy effects of the long-range
Coulomb interaction in $1D$ electron systems, in order to calculate
the effects of a strong transverse magnetic field.
 Our main
focus has been the scaling behavior of quantities like the
quasiparticle weight or the low-energy DOS, which
can be compared directly with the results of transport
experiments.
 For this
purpose, we have developed two RG approaches: the first one at
dimension $D$ strictly equal to $1$; the second one  in a
dimensionally regularized theory devised to interpolate between
$D=2$ and $D=1$.


One of the most significant observations made in MWNTs has
been the power-law behavior of the tunneling conductance as a
function of the temperature or the bias voltage.  The measurements
carried out in MWNTs have displayed a power-law behavior of
the tunneling conductance, that gives a measure of the low-energy
DOS, with exponents ranging from $0.24$ to $0.37$
\cite{22n}. These values are, on the average, below those measured
in SWNTs, which are typically about $0.35$
\cite{12n}. In recent papers\cite{npb} we showed that our
results can account satisfactorily for this slight reduction in
the critical exponent with the change of the nanotube thickness.

In a recent letter Kanda et al.\cite{kprl} examined the dependence
of $G$ on perpendicular magnetic fields in MWNTs. They found that
the exponent $\alpha$  depends significantly on the gate voltage,
giving strong oscillations. However the value of $G$ depends not
only on the gate voltage, but also on the magnetic field and, in most cases, $G$
is smaller for higher magnetic fields.
 These authors showed that $\alpha$ is reduced from a value of
$0.34$ to a value $0.11$ for a magnetic field ranging from $0$ to
$4$ T, in correspondence of a peak in the $\alpha$ oscillations
(and from a value of $0.06$ to a value $0.005$ for a different
value of the gate voltage $V_g$, which corresponds to a dip in the
$V_g$ dependence of the conductance).

\

Now we can compare our prediction with the experimental values
starting from the strictly $1D$ approach. In order to do that, we
start from eq.(\ref{az0}) by introducing the potential in
eq.(\ref{vq}).  It follows that a realistic value of the effective
coupling at vanishing magnetic field is $U_0(q_c,0)/(2\pi
v_F)\approx 8$ which gives a value for the critical exponent
$\alpha \approx 0.33$. Our calculation predicts that there should
be a significant reduction of $\alpha$ as the magnetic field
increases for CNs of large radius. We just recall that a field
$B$ of $4 T$ corresponds to a value of $\gamma \approx 1$, so that
 the reduction of the interaction strength caused by the
growth of the magnetic field gives a value of
$U_0(q_c,\omega_c)/(2\pi v_F)\approx 8/3$, which corresponds to
$\alpha=0.11$, in agreement with the experimental results.

Kanda et al\cite{kprl} also found a reduction at $B=0$  in the
effective interaction acting on the gate voltage. In order to
explain this result, we can introduce a different value of
$U_0(q_c,0)/(2\pi v_F)\approx 1.42$, corresponding to the measured
$\alpha=0.06$. This effect could be due to the effect of the
single particle spectrum\cite{conmat} and  to the reduction of the
Fermi velocity corresponding to a shift of the Fermi level.
 In this case a reduction in the coupling $U_0(q_c,\omega_c)$ of a factor $1/3$,
 due to the magnetic field, gives a strongest reduction in the critical
 exponent with $\alpha$ below $0.01$, in agreement with the
 measurements.

\
In previous papers\cite{npb} we showed that the dimensional
crossover approach  is appropriate for the description of CNs of
large radius. There we found a value of the critical exponent in
agreement with the experimental data.

As we discussed previously, the rescaling of all repulsive
terms of the interaction between electrons is strongly  due to
 the edge localization of the electrons, corresponding to
 a behaviour which is strictly 1D.
This effect yields the result that the Luttinger liquid critical dimension
(corresponding to the crossover dimension) is strongly reduced,
when the magnetic field increases. A simple mechanism can explain
this behavior. When $B$ increases, because of the localization of the
edge states, at any dimension greater than 1 the CN is more
similar to a 2D graphene sheet than to a 1D
wire. Only when the energy goes below a small value, the system
restores its Luttinger liquid behavior and the quasiparticle
amplitude vanishes, following the usual behavior.

Hence, we predict the disappearance of the Luttinger Liquid at
very strong magnetic field, and we can reduce this phenomenon to
the strong localization of the edge states which imposes a 2D
behaviour to the system.

It is clear from this picture that the main effect of a strong
magnetic field is the fast renormalization of the coupling, which
vanishes as the dimension is different from $D=1$. It follows that,
for a strong magnetic field, the quasiparticle weight is not
renormalized to $Z=0$  and the Luttinger liquid disappears. In
this case our approach fails, because the forward scattering
between currents with different chirality vanishes, coherently with
the assumed formation of a chiral liquid, where obviously
$\alpha$ is zero.

The main prediction that comes from our study is that there should
be a significant reduction in the critical exponent of the
tunneling DOS, as the transverse magnetic field  is
increased in nanotubes of large radius. It would be relevant to
test such a dependence in experiments carried out at various values of the
magnetic field, using different samples.
\appendix

\section{From the 2D Coulomb potential to a 1D Model}

{We approximate the Coulomb potential function, in the limit of the small ratio
$R/|y-y'|$, as
$$
U({\bf r}-{\bf r'})=\frac{c_0}{|y-y'|}\left(\sum_k^\infty
\frac{{\left( -1 \right) }^k\,\,\Gamma(\frac{1}{2} + k)}
  {{\sqrt{\pi }}\,\Gamma(1 + k)} \left(\frac{2R}{y-y'}\right)^{2\,k}\sin^{2\,k}(\frac{\varphi-\varphi'}{2})\right).
$$

The Forward scattering between opposite branches is obtained as
\bea \nonumber
 U(y-y')&=& c_0 \int_{-\pi}^\pi
d\varphi\int_{-\pi}^\pi d\varphi'\sqrt{\frac{1}{(y-y')^2+4 R^2
\sin^2(\frac{\varphi-\varphi'}{2})}}\widetilde{\Phi^*}_{0,k_F}(\varphi,y)\widetilde{\Phi}_{0,k_F}(\varphi,y)\widetilde{\Phi}_{0,-k_F}(\varphi',y')\widetilde{\Phi^*}_{0,-k_F}(\varphi',y')
\\&=&\frac{c_0}{4\pi^2}\int_{-\pi}^\pi
d\varphi\int_{-\pi}^\pi d\varphi'\sqrt{\frac{1}{(y-y')^2+4 R^2
\sin^2(\frac{\varphi-\varphi'}{2})}} \left(\frac{(1-2\,{\gamma
}\,\cos(\varphi' ))^2}{{\left( 1 + 2 \,{\gamma }^2
        \right) }}\right)\left(\frac{(1+2\,{\gamma }\,\cos(\varphi' ))^2}{{\left( 1 +
2 \,{\gamma }^2
        \right) }}\right)
\nonumber \\
       &=&\frac{c_0}{4\pi^2}\sqrt{\frac{1}{(y-y')^2}}\int_{-\pi}^\pi
d\varphi\int_{-\pi}^\pi d\varphi'\left(\sum_k^\infty \frac{{\left(
-1 \right) }^k\,\,\Gamma(\frac{1}{2} + k)}
  {{\sqrt{\pi }}\,\Gamma(1 + k)} \left(\frac{2R}{y-y'}\right)^{2\,k}\sin^{2\,k}(\frac{\varphi-\varphi'}{2})\right)
\nonumber
  \\ &\times&
  \left(\frac{(1-4\,\,{\gamma }^2\,\cos(\varphi )\,\cos(\varphi' ) +\,{\gamma }^2\,(\cos^2(\varphi )+\cos^2(\varphi' )))}{{\left(
1 + 2 \,{\gamma }^2
        \right)^2 }}\right)
\nonumber \\
        & =&
        8 c_0\sqrt{\frac{\pi^3}{(y-y')^2}}\sum_k^\infty \frac{{\left(
-1 \right) }^k\,\,\Gamma(\frac{1}{2} + k)}
  {{\sqrt{\pi }}\,\Gamma(1 + k)} \left(\frac{2R}{y-y'}\right)^{2\,k}\nonumber \\
  &\times& \left( \frac{\Gamma(n+1/2)}{\Gamma(n+1)}-2\,{\gamma }^2 \frac{n \Gamma(n+1/2)}{\Gamma(n+1)}
  +\,{\gamma }^2\left[\frac{\Gamma(n+1/2)}{\Gamma(n+2)}+\frac{2\Gamma(n+3/2)}{\Gamma(n+2)} \right]
  \right).
 \eea
Hence, we obtain
 \bea
 U(y-y')&=& 2 \frac{c_0}{\left( 1 + 2 \,{\gamma }^2
        \right)^2}\sqrt{\frac{1}{(y-y')^2}}\nonumber \\
        &\times& \left\{K(-(\frac{2R}{y-y'})^2) +\frac{\pi \gamma^2 }{4}
 \left( 4 _2F_1(\frac{1}{2},\frac{3}{2};2,-(\frac{2R}{y-y'})^2)+(\frac{2R}{y-y'})^2_2F_1(\frac{3}{2},\frac{3}{2};2,-(\frac{2R}{y-y'})^2)\right)           \right\}
 \eea
 where  $K_E(x)$ gives the complete elliptic integral of the first
kind, while $_2F_1(a,b,c,z)$  is the hypergeometric function.

The Fourier transform gives $U_0(q)$ as \bea
U_0(q)=\frac{c_0}{\sqrt{2}\left( 1 + 2 \,{\gamma }^2
        \right)^2}\left[K_0(\frac{q}{2})I_0(\frac{q}{2}) +{\gamma }^2 \left(2G_1(\frac{q^2}{4})+G_2(\frac{q^2}{4})\right)\right],
\eea with $K_n(q)$ denoting the modified Bessel function of
the second kind and $I_n(q)$ corresponding to the modified Bessel
function of the first kind. Here we have $G_1(z)={MeijerG}(\{ \{ -\left(
\frac{1}{2} \right)
       \} ,\{ \} \} ,\{ \{ 0,0\} ,\{ -1\} \} ,
  z)$ and $G_2(z)={MeijerG}(\{ \{ \frac{1}{2}\} ,\{ \} \} ,
  \{ \{ 0,1\} ,\{ 0\} \} ,\frac{q^2}{4})$, in terms of $MeijerG$ functions. }
   The Backward scattering is obtained analogously and we obtain
\bea
 U(y-y')&=& 2 c_0\sqrt{\frac{1}{(y-y')^2}} \left\{K(-(\frac{2R}{y-y'})^2) -4\pi \gamma^2
\, _2F_1(\frac{1}{2},\frac{3}{2};2,-(\frac{2R}{y-y'})^2) \right\}.
 \eea

The Fourier transform gives the $U_0(2k_F)$ as

\bea U_0(2k_F)=\frac{c_0}{\sqrt{2}\left( 1 + 2 \,{\gamma }^2
        \right)^2}\left[K_0(k_F)I_0(k_F) -8{\gamma }^2 G_1(k_F^2)\right].
\eea

\section{Coulomb interaction in general dimension}
We can start from the general Coulomb interaction in 3D and
remember that the length of the system in one direction (e.g. the $x$ one) is
quite smaller than in the other ($L_y$)
 \bea
 U({\bf r}-{\bf r'})
 \approx   \frac{c_0}{4\pi^2|{\bf r}-{\bf r'}|}\left(\sqrt{\frac{L_y^2}{L_y^2+(x-x')^2)}} \right)=
        \frac{c_0}{|{\bf r}-{\bf r'}|}\Upsilon_2(\omega_c),
 \eea
 where ${\bf r}$ is a $2D$ vector and
$$
\Upsilon_{k,p,q}(\omega_c)=\int \int dx dx'
\left(\sqrt{\frac{L_y^2}{L_y^2+(x-x')^2)}} \right) u_{0}\left(x,
k\right)u_{0}\left(x,p\right) u_{0}\left(x', (k+q)\right).
u_{0}\left(x',(p-q)\right)
$$
In the limit of Forward scattering we obtain \bea
\Upsilon_{F}(\omega_c)&=&\frac{1}{\left( 1 + \,{\gamma }^2
        \right)^2}\nonumber \\
        &\times& \left\{K(-(\frac{2R}{L_y})^2) +\frac{\pi \gamma^2 }{4}
 \left( 4 _2F_1(\frac{1}{2},\frac{3}{2};2,-(\frac{2R}{L_y})^2)+(\frac{2R}{L_y})^2_2F_1(\frac{3}{2},\frac{3}{2};2,-(\frac{2R}{L_y})^2)\right) \right\}.
\eea

\bibliographystyle{prsty} 
\bibliography{}

\end{document}